\documentclass[aps,prb,twocolumn,superscriptaddress,showpacs,10pt]{revtex4-1}

\usepackage{bm}
\usepackage{graphicx}
\usepackage{epsfig}
\usepackage{longtable}
\usepackage{amsmath}
\usepackage{amssymb}
\usepackage{color}

\usepackage{hyperref}

\begin{document}

\title{Experimental determination of Rashba spin-orbit coupling in wurtzite $n$-GaN:Si}

\author{W. Stefanowicz}
\affiliation{Institute of Physics, Polish Academy of Sciences, al. Lotnik\'ow 32/46, PL-02 668 Warszawa, Poland}

\author{R. Adhikari}
\affiliation{Institut f\"ur Halbleiter-und-Festk\"orperphysik, Johannes Kepler University, Altenbergerstr. 69, A-4040 Linz, Austria}

\author{T. Andrearczyk}
\affiliation{Institute of Physics, Polish Academy of Sciences, al. Lotnik\'ow 32/46, PL-02 668 Warszawa, Poland}

\author{B. Faina}
\affiliation{Institut f\"ur Halbleiter-und-Festk\"orperphysik, Johannes Kepler University, Altenbergerstr. 69, A-4040 Linz, Austria}

\author{M.~Sawicki}
\affiliation{Institute of Physics, Polish Academy of Sciences, al. Lotnik\'ow 32/46, PL-02 668 Warszawa, Poland}

\author{J. A. Majewski}
\affiliation{Institute of Theoretical Physics, Faculty of Physics, University of Warsaw, ul. Ho\.za 69, PL-00 681 Warszawa, Poland}

\author{T. Dietl}
\email{mailto: dietl@ifpan.edu.pl}
\affiliation{Institute of Physics, Polish Academy of Sciences, al. Lotnik\'ow 32/46, PL-02 668 Warszawa, Poland}
\affiliation{Institute of Theoretical Physics, Faculty of Physics, University of Warsaw, ul. Ho\.za 69, PL-00 681 Warszawa, Poland}
\affiliation{WPI-Advanced Institute for Materials Research (WPI-AIMR), Tohoku University, 2-1-1 Katahira, Aoba-ku, Sendai 980-8577, Japan}

\author{A. Bonanni}
\email{mailto: alberta.bonanni@jku.at}
\affiliation{Institut f\"ur Halbleiter-und-Festk\"orperphysik, Johannes Kepler University, Altenbergerstr. 69, A-4040 Linz, Austria}

\begin{abstract}
Millikelvin magnetotransport studies are carried out on heavily $n$-doped wurtzite GaN:Si films grown on semi-insulating GaN:Mn buffer layers by metal-organic vapor phase epitaxy. The dependence of the conductivity on magnetic field and temperature is interpreted in terms of theories that take into account disorder-induced quantum interference of one-electron and many-electron self-crossing trajectories. The Rashba parameter $\alpha_{\text{R}}\,=\,(4.5 \pm 1)$\,meV{\AA} is determined, and it is shown that in the previous studies of electrons adjacent to GaN/(Al,Ga)N interfaces, bulk inversion asymmetry was dominant over structural inversion asymmetry. The comparison of experimental and theoretical values of $\alpha_{\text{R}}$ across a series of wurtzite semiconductors is presented as a test of current relativistic {\em ab initio} computation schemes. Low temperature decoherence is discussed in terms of disorder-modified electron-electron scattering.
\end{abstract}

\date{\today}


\maketitle

\section{Introduction}
Beside being strategic materials systems for nowadays optoelectronic\cite{Morkoc:2009_B} and high-power applications,\cite{Morkoc:2009_B,Palacios:2005_IEEE} GaN and related alloys are expected to play a major role in the realization of spin-related functionalities based on semiconductors. In particular, Fe doping serves routinely to fabricate semi-insulating GaN substrates.\cite{Kaun:2013_SST} However, GaN with higher Fe concentrations exhibits room temperature ferromagnetic\cite{Bonanni:2008_PRL,Navarro:2010_PRB} and antiferromagnetic\cite{Navarro:2010_PRB} features associated with the aggregation of Fe cations, leading to the formation of various magnetically robust Fe$_x$N nanocrystals. Furthermore, light absorption associated with the Mn mid-gap band in GaN improves the efficiency of GaN-based solar cells.\cite{Sheu:2013_APL} At the same time, the formation of Mn-Mg$_k$ impurity complexes activates room temperature infrared luminescence,\cite{Devillers:2012_SR} suggesting that the optoelectronic capabilities of nitrides can be extended towards the communication windows.

Other appealing aspects of these systems are associated with spin-orbit coupling (SOC). On the one hand, the small value of the nitrogen proton number $Z_{\text{a}}$, leading to weak spin-orbit splitting $\Delta_{\text{so}}$ of the valence band, results in a long spin relaxation time of electrons in GaN.\cite{Beschoten:2001_PRB,Krishnamurthy:2003_APL} On the other hand, strong interfacial electric fields result in substantial Rashba-type SOC that in the extreme case of GaN/InN/GaN quantum wells may lead to a transition to the topological insulator phase.\cite{Miao:2012_PRL} In this context particularly appealing is the demonstration that (Ga,Mn)N is a dilute ferromagnetic insulator,\cite{Bonanni:2011_PRB,Sawicki:2012_PRB,Stefanowicz:2013_PRB} which offers prospects to the search for phenomena associated with the interplay between SOC and the exchange splitting of bands by ferromagnetic proximity effects.\cite{Sau:2010_PRL}

Here, we report on experimental studies of weak localization and antilocalization magnetoresistance in epitaxial layers of wurtzite (wz) $n$-GaN:Si. A theoretical description of the data in terms of the Hikami, Larkin, and Nagaoka theory,\cite{Hikami:1980_PTP} suitably adapted for wz compounds,\cite{Sawicki:1986_PRL} allows us to extract the magnitude of the parameter $\alpha_{\text{R}}$ describing the Rashba term linear in $k$ and accounting for the spin-splitting of the conduction band in bulk wz semiconductors.\cite{Rashba:1960_SPSS} This term is specific to the crystal structure in question and its presence is {\em not} associated with interfacial electric fields.\cite{Bychkov:1984_JPC}
The value of $\alpha_{\text{R}}=(4.5\pm1$)\,meV{\AA} we determine here is by two orders of magnitude greater
than the one found by electron spin resonance (ESR) for electrons trapped by donors or accumulated at the surface of $n$-GaN.\cite{Wolos:2011_PB} At the same time, it agrees with the value found for electrons attracted to the GaN/(Al,Ga)N interface by polarization electric fields.\cite{Schmult:2006_PRB, Kurdak:2006_PRB,Thillosen:2006_APL, Thillosen:2006_PRB,Zhou:2008_JAP,Cheng:2008_PE,Belyaev:2008_PRB} Our results demonstrate, therefore, that the bulk rather than the structure inversion asymmetry accounts for the spin splitting of interfacial states.  We discuss also the chemical trends in $\alpha_{\text{R}}$ and show that the discrepancy between the current  {\em ab initio} theories\cite{Voon:1996_PRB,Majewski:2005_P} and the present and previous magnetoresistance,\cite{Sawicki:1986_PRL,Andrearczyk:2005_PRB} optical\cite{Romestain:1977_PRL,Dobrowolska:1984_PRB} and ESR studies\cite{Kozuka:2013_PRB} is within a factor of two for various wz-$n$-type semiconductors: ZnO, GaN, CdS, and CdSe. Finally, we treat the temperature dependence of the phase coherence length $L_{\varphi}$ and conductivity $\sigma$ in terms of electron-electron interactions in disordered systems.\cite{Altshuler:1985_B,Lee:1985_RMP,Altshuler:1982_JPC}

\section{samples and experiment}

The Si-doped GaN layers considered in the present study have been grown in an AIXTRON 200RF horizontal tube metalorganic vapor phase epitaxy (MOVPE) reactor and deposited on a $c$-plane sapphire substrate using TMGa, MnCp$_{2}$, NH$_{3}$, and SiH$_{4}$ as precursors for Ga, Mn, N, and Si respectively, with H$_{2}$ as carrier gas. After nitridation of the sapphire substrate, a low temperature nucleation layer (NL) is deposited at 540$^{\circ}$C and then annealed at 1040$^{\circ}$C. Successively, a 1\,$\mathrm{\mu}$m-thick GaN:Mn buffer layer is grown also at 1040$^{\circ}$C, Mn being introduced in order to compensate the $n$-type background proper of the GaN layers fabricated by MOVPE. The concentration of Mn in the buffer layer is as low as 0.06\%, as confirmed by secondary-ion mass spectroscopy (SIMS) and SQUID magnetometry. A 150\,nm layer of GaN:Si is further grown at 1000$^{\circ}$C onto the GaN:Mn buffer layer. All steps of the growth process are monitored with \textit{in situ} spectroscopic and kinetic ellipsometry.

The grown samples are systematically characterized by atomic force microscope (AFM), high resolution x-ray diffraction (HRXRD), high resolution transmission electron microscopy (HRTEM) and SIMS has been employed for chemical analysis. The AFM micrographs reveal a flat surface (rms roughness $\approx$1\,nm) while HRXRD and HRTEM confirm the high crystallinity of the samples.
The HRTEM analysis does not reveal any secondary phases like $e.\,g.$ precipitates of Si$_\mathrm{x}$N, and energy dispersive x-ray spectroscopy (EDS) states the homogeneous distribution of Si in the doped layer with a concentration of 0.24\%, in agreement with SIMS data.

The magnetotransport measurements have been performed on Hall bars with Ti/Au/Al/Ti/Au metallic contacts fabricated by conventional photolithography. Transport experiments down to 40\,mK have been carried out in a dedicated home-built helium cryostat and a dilution refrigerator. The electron concentration obtained from the Hall data is found to be constant over a wide range of temperatures and to have a value $n=1.2\times 10^{19}$\,cm$^{-3}$, far over the critical value for the metal-to-insulator (MIT) transition in bulk GaN, $n_\mathrm{MIT}\,\approx\,10^{18}$\,cm$^{-3}$.\cite{daSilva:2002_JAP}
The degenerate and metallic character of the samples is further documented by the absence of dependence of $\sigma$ on the temperature in the limit $T\,\rightarrow\,0$, as well as by the values of the Hall mobility $\mu = 140$\,cm$^2$/Vs and $k_{\text{F}}\ell = 4.6$ in this regime, where $\ell = \hbar k_{\text{F}}\mu/e$ is the mean free path. Accordingly, we interpret the measured magnetoresistance $\Delta \rho(T,B)$ in terms of quantum corrections to the conductivity of disordered systems, developed for $k_{\text{F}}\ell > 1$ and $\ell < l_B = (\hbar/eB)^{1/2}$.\cite{Altshuler:1985_B,Lee:1985_RMP}

\section{experimental results}

\begin{figure}[ht]
	\centering
	\includegraphics[width=0.75\linewidth]{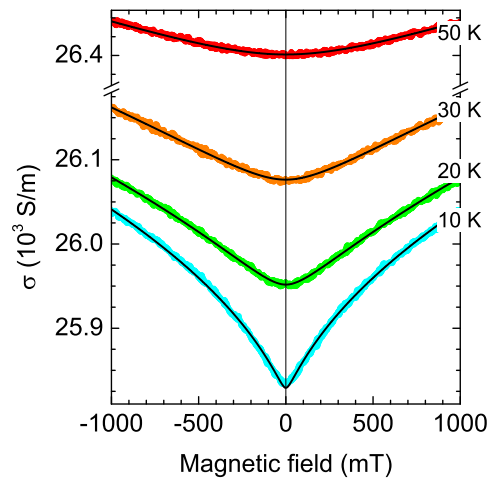}
	\caption{(Color online) Magnetoconductivity at $T \geq 10$\,K (points: experimental data; solid lines: theoretical fitting). The 3D theory of conductivity changes in the magnetic field by Kawabata\cite{Kawabata:1980_JPSJ,Kawabata:1980_SSC} is employed with the phase coherence length $L_{\varphi}(T)$ as the only fitting parameter.}
	\label{fig:fig1}	
\end{figure}

\begin{figure}[ht]
	\centering
	\includegraphics[width=0.75\linewidth]{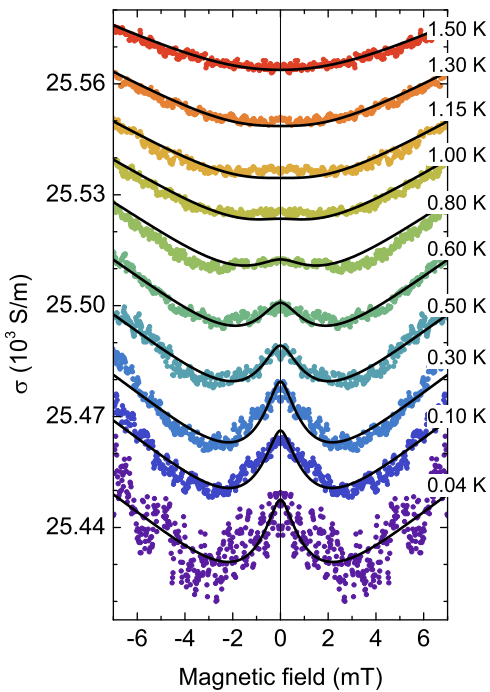}
	\caption{(Color online) Magnetoconductivity at  $T\leq 1.5$\,K (points: experimental data; solid lines: theoretical fitting). The fitting is performed within the 2D model of conductivity changes in the magnetic field by Hikami {\em et al.},\cite{Hikami:1980_PTP} treating $\alpha_{\text{R}}$ and $L_{\varphi}(T)$ as fitting parameters. The two lowest curves are down-shifted for clarity.}
	\label{fig:fig2}	
\end{figure}

In Figs.\,\ref{fig:fig1} and \ref{fig:fig2} the conductivity $\sigma(T,B) =1/\rho(T,B)$ of the GaN:Si film is shown at different temperatures $T$ as a function of the magnetic field $B$ applied perpendicular to the film surface, $i.\,e.$ parallel to the wz-$c$-axis. It is seen from Fig.\,\ref{fig:fig1} that for $T \geq 10$\,K the magnetoconductivity (MC) is solely positive, while from the data collected in Fig.\,\ref{fig:fig2} for $T \leq 1.5$\,K and down to 40\,mK there are contributions of both negative and positive MC in low magnetic fields. This negative component of MC is related to the appearance of a weak antilocalization (WAL) maximum (which vanishes above 1\,K), a distinct signature of SOC.

For $T \geq 10$\,K, the experimental results are fitted within a three dimensional (3D) theoretical model of weak localization MC in semiconductors on the metallic side of the MIT, as proposed by Kawabata.\cite{Kawabata:1980_JPSJ} Here, the phase coherence length $L_{\varphi}(T)$ is the only fitting parameter. As seen, the theory describes the data quite accurately. We add that the presence of positive MC is often taken as an indication for spin disorder scattering. Actually, even in the presence of magnetic impurities, spin disorder scattering is typically masked by other scattering mechanisms in semiconductors, and rarely perturbs the conductivity directly.

However, this 3D model does not describe the observed MC for $T \leq 5$\,K and low magnetic fields, where two additional aspects must be considered, namely: (i) the impact of SOC on the quantum corrections to the conductivity, leading to a WAL maximum in MC below 1\,K (as seen in Fig.\,\ref{fig:fig2}) and (ii) a dimensional crossover from 3D to 2D that occurs if $L_\varphi(T) \gtrsim d$, where $d$ is the layer thickness. In Fig.\,\ref{fig:fig2}, the fingerprint of WAL is observed for $T\lesssim 1$\,K and for $B \lesssim 3$\,mT. The MC data obtained for $T \leq 1.5$\,K are fitted with the theoretical model proposed  for 2D films in the weakly localized regime, $k_{\text{F}}\ell> 1$, and considering effects of spin-dependent scattering.\cite{Hikami:1980_PTP}

According to the $\mathbf{k}\cdot\mathbf{p}$ theory, SOC in wz semiconductors leads to a term linear in $k$ in the effective mass equation,\cite{Rashba:1960_SPSS,Casella:1960_PRL}
\begin{equation}
{\cal{H}}_{\text{so}} = \alpha_{\text{R}}\hat{c}\cdot(\vec{\sigma}\times\vec{k}),
\end{equation}
where $\alpha_{\text{R}}$ is the Rashba $\mathbf{k}\cdot\mathbf{p}$ parameter; $\hat{c}$ is the versor along the wz-$c$-axis, and $\vec{\sigma}$ are the Pauli matrices. For $\alpha_{\text{R}}k_{\text{F}}\tau/\hbar <1$,
the corresponding spin relaxation times are given by,\cite{Altshuler:1982_B}
\begin{equation}
\tau_{\text{sox}}^{-1} = \tau_{\text{soy}}^{-1} = \alpha_{\text{R}}^2k_{\text{F}}^2\tau/3\hbar^2;\; \tau_{\text{soz}}^{-1} =0,
\end{equation}
where the $z$ axis is taken along the wz-$c$-axis, $\tau=\mu m^*/e$ is the momentum relaxation time, and he effective mass is $m^* =0.22m_0$ for GaN.\cite{Witowski:1999_APL}

Here, two fitting parameters $\alpha_{\text{R}}$ and $L_{\varphi}(T)$, are employed to describe the conductivity changes in magnetic field. From the fitting of the MC data in the low temperature range, we find $\alpha_{\text{R}}$ for wz-$n$-GaN:Si to be $(4.5\pm1)$\,meV{\AA}. Within the experimental uncertainties, this value is virtually identical to the one determined by the interfacial polarization electric field ${\cal{E}}$\ in numerous MC studies of a 2D electron gas adjacent to the GaN/(Al,Ga)N interface.\cite{Schmult:2006_PRB,Kurdak:2006_PRB,Thillosen:2006_APL,Thillosen:2006_PRB,Zhou:2008_JAP,Cheng:2008_PE}  This agreement means that the Rashba spin relaxation associated with the wz crystal structure dominates over effects brought about by the interfacial field ${\cal{E}}$. This also explains why the value of $\alpha_{\text{R}}$ determined for GaN/(Al,Ga)N heterostructures was found to be independent of the gate electric field and of the Al content in the barrier.\cite{Thillosen:2006_PRB}

\begin{figure}[ht]
	\centering
	\includegraphics[width=0.75\linewidth]{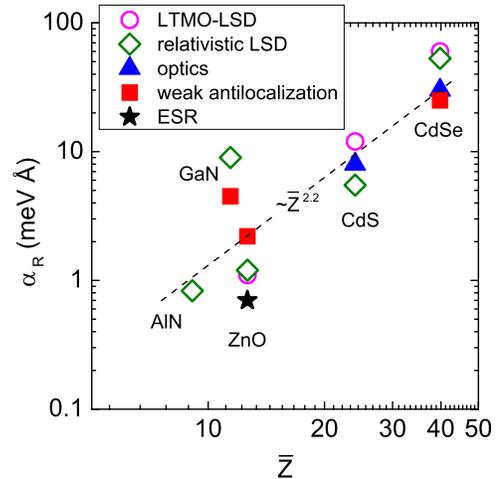}
	\caption{(Color online) Rashba parameter $\alpha{_\text{R}}$ as a function of the harmonic average proton number $\bar{Z}$, as determined experimentally from (i) weak antilocalization (solid squares) for GaN:Si (this work), ZnO:Al,\cite{Andrearczyk:2005_PRB}  and CdSe:In;\cite{Sawicki:1986_PRL}  (ii) optical studies (solid triangles) for $n$-CdS (Ref.~\onlinecite{Romestain:1977_PRL}) and $n$-(Cd,Mn)Se (Ref.~\onlinecite{Dobrowolska:1984_PRB}), and (iii) electron spin resonance (star) for $n$-ZnO/(Zn,Mg)O,\cite{Kozuka:2013_PRB} compared to results of {\em ab initio} computations [open circles (Ref.\,\onlinecite{Voon:1996_PRB}) and open diamonds (Ref.\,\onlinecite{Majewski:2005_P} and this work)]. Dashed line shows  $\bar{Z}^{2.2}$ dependence.}
	\label{fig:fig3}
\end{figure}

In Fig.\,\ref{fig:fig3} we provide a compilation of $\alpha_{\text{R}}$ values determined through various experimental methods for wz semiconductor compounds, plotted as a function of a harmonic average of the cation and anion proton numbers, $\bar{Z} = 2 (1/Z_{\text{c}} + 1/Z_{\text{a}})^{-1}$, and compared to results of {\em ab initio} computations in the framework of the density functional theory with relativistic effects taken into account non-perturbatively.\cite{Majewski:2004_PSSc} A chemical trend, $\alpha_{\text{R}} \sim \bar{Z}^{2.2 \pm 0.5}$, is evident and confirms that the significance of the SOC increases with the square of the nucleus charge. Furthermore, it is seen that the theory describes the experimental values with an accuracy better than a factor of two.

\begin{figure}[ht]
	\centering
	\includegraphics[width=0.75\linewidth]{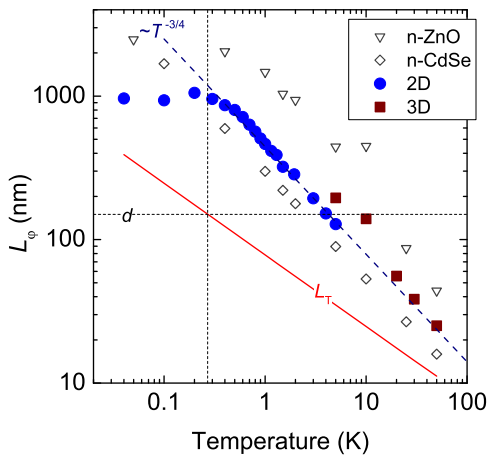}
	\caption{(Color online) Phase coherence length $L_\varphi(T)$ obtained by fitting the magnetoconductance data for $n$-GaN (Figs.\,1 and 2) within the 3D and 2D models (squares and circles, respectively). For comparison the corresponding data for a film of $n$-ZnO (Ref.~\onlinecite{Andrearczyk:2005_PRB}) and a bulk sample of $n$-CdSe (Ref.~\onlinecite{Sawicki:1986_PRL}) are also presented. Dashed line: dependence $T^{-3/4}$; solid line: $L_T$ for $n$-GaN marking a cross-over from 3D to 2D at $L_T \simeq d$, where $d$ is the film thickness.}
	\label{fig:fig4}
\end{figure}

Our fitting procedure provides also the values of $L_{\varphi}(T)$, which are shown in Fig.~\ref{fig:fig4}, and compared to corresponding data for CdSe:In (Ref.~\onlinecite{Sawicki:1986_PRL}) and ZnO:Al\,(Ref.\,\onlinecite{Andrearczyk:2005_PRB}). A dependence $L_{\varphi}(T)  \propto T^{-3/4}$ is observed for all compounds over a wide temperature range, a behavior expected theoretically for decoherence brought about by electron-electron interactions in disordered 3D systems.\cite{Altshuler:1982_JPC,Altshuler:1985_B} A transition to the 2D case occurs in this case at $L_T = \hbar(k_{\text{F}}\ell /3k_{\text{B}}Tm^*)^{1/2} \gtrsim d$. A change in the $L_{\varphi}(T)$ slope observed in this region, if not caused by noise-related decoherence, can be associated with the dimensional cross-over.

According to theoretical expectations, $L_{\varphi}(T)$ has the same functional form for electron-electron scattering with large and small energy transfers in the 3D case.\cite{Altshuler:1982_JPC}  The relaxation times corresponding to these two contributions are given by,\cite{Altshuler:1982_JPC,Altshuler:1985_B}
\begin{eqnarray}
\tau_{\varphi1} &=& 12\sqrt{2}\pi^2  \hbar \nu_F \biggl[1+\frac{3F}{4+F}\biggl(\bigl(1+\frac{F}{2}\bigr)^{3/2}-1\biggr)\biggr]^{-1} L_T^{3}; \nonumber \\
\tau_{\varphi2} &=& \frac{8}{3}\frac{k_F^2 \ell L_T^3}{D}.
\end{eqnarray}
where the energy $\epsilon$ of quasiparticles is identified with $k_BT$,  $\nu_F = m^* k_F/(\pi^2 \hbar^2)$ is the density of states at the Fermi level, and $F$ is the coupling parameter in the triplet channel. This parameter controls also the temperature dependence of the conductivity at low temperatures, whose studies lead to $F = 2.73$ for CdSe:In (Ref.\,\onlinecite{Sawicki:1986_PRL}) and $0.15$ for GaN:Si, as discussed below.

By taking $\tau_{\varphi}=(1/\tau_{\varphi1}+1/\tau_{\varphi2})^{-1}$ and $L_\phi=(D \tau_\varphi)^{1/2}$, we evaluate $a= 550$ and 1670\,nm\,K$^{3/4}$ for the sample of CdSe:In and GaN:Si, respectively, where $a$ is the prefactor in the dependence $L_{\varphi}(T) = aT^{-3/4}$. Furthermore, our estimation confirms that scattering with large energy transfers dominates, $1/\tau_{\varphi1} \gg 1/\tau_{\varphi2}$.
However, the theoretical values of $a$ are for both materials greater than those implied by the experimental data presented in Fig.~\ref{fig:fig4},  for which $a = 300$ and 440\,nm\,K$^{3/4}$, respectively.

\begin{figure}[ht]
	\centering
	\includegraphics[width=0.7\linewidth]{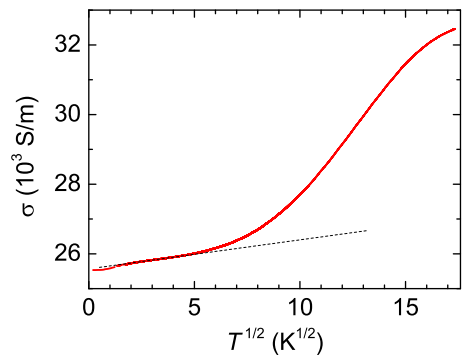}
	\caption{(Color online) Dependence of the conductivity on square root temperature. Dashed line: linear fit for temperatures $\leq 25$\,K, this behavior being explained in terms of disorder-modified electron-electron interactions.}
	\label{fig:fig5}	
\end{figure}

In Fig.\,\ref{fig:fig5} the zero magnetic field conductivity $\sigma(T)$ is reported as a function of $T^{1/2}$, and it is seen to have a linear dependence on square root of temperature below 25\,K. This behavior is assigned to quantum corrections to the conductivity due to disorder-modified electron-electron interactions.\cite{Altshuler:1985_B, Lee:1985_RMP} A quantitative comparison of this slope to the theory leads to a value of the coupling parameter in the triplet channel $F=0.15$. This relatively small value can be expected for $k_{\text{F}}\ell  \gg 1$. However, on approaching the metal-to-insulator transition the value of $F$ increases. This effect was observed in $n$-CdSe, where $F\approx{2.7}$ and the conductivity decreases with increasing temperature.\cite{Sawicki:1986_PRL}

\section{conclusions}

In summary, we have carried out low temperature magnetotransport studies on high quality heavily doped wz-GaN:Si films grown on a semi-insulating GaN:Mn buffer layer. Our investigations reconfirm the relevance, in doped degenerate semiconductors, of disorder-induced quantum interference of one-electron and many-electron self-crossing trajectories, effects not captured by the Drude-Boltzmann description of transport phenomena. The quantitative models of the magnetoconductance data have allowed us to determine the Rashba parameter $\alpha_{\text{R}}$, so far known only from studies of 2DEG adjacent to GaN/(Al,Ga)N interfaces. Our results demonstrate that inversion asymmetry associated with the wurtzite crystal structure dominates over the effects of the interfacial electric field in the conduction band of GaN. The comparison of experimental and theoretical values of $\alpha_{\text{R}}$ across a series of wz semiconductors has provided an important test of the current relativistic {\em ab initio} computation schemes, demonstrating that the differences between the experimental and theoretical values are within a factor of two.  Furthermore, our quantitative interpretation of the decoherence length $L_{\varphi}(T)$ has confirmed that electron-electron scattering is a dominant source of phase breaking in these systems. With these premises, wide perspectives open for nitrides, as building-blocks for the next generation of spin devices exploiting spin-orbit coupling and the magnetism of transition-metal doped layers.

\section*{acknowledgments}

This work was supported by the FunDMS Advanced Grant of the European Research Council (ERC Grant No.\,227690) within the Ideas 7$^{\mathrm{th}}$ Framework Programme of the European Union,
and by the Austrian Science Foundation -- FWF (P20065, P22477, and P24471),
and by National Science Centre (Poland) under OPUS grants No.\,2011/03/B/ST3/02457 and  2013/09/B/ST3/04175.


\begin{thebibliography}{0}%
\makeatletter
\providecommand \@ifxundefined [1]{%
 \@ifx{#1\undefined}
}%
\providecommand \@ifnum [1]{%
 \ifnum #1\expandafter \@firstoftwo
 \else \expandafter \@secondoftwo
 \fi
}%
\providecommand \@ifx [1]{%
 \ifx #1\expandafter \@firstoftwo
 \else \expandafter \@secondoftwo
 \fi
}%
\providecommand \natexlab [1]{#1}%
\providecommand \enquote  [1]{``#1''}%
\providecommand \bibnamefont  [1]{#1}%
\providecommand \bibfnamefont [1]{#1}%
\providecommand \citenamefont [1]{#1}%
\providecommand \href@noop [0]{\@secondoftwo}%
\providecommand \href [0]{\begingroup \@sanitize@url \@href}%
\providecommand \@href[1]{\@@startlink{#1}\@@href}%
\providecommand \@@href[1]{\endgroup#1\@@endlink}%
\providecommand \@sanitize@url [0]{\catcode `\\12\catcode `\$12\catcode
  `\&12\catcode `\#12\catcode `\^12\catcode `\_12\catcode `\%12\relax}%
\providecommand \@@startlink[1]{}%
\providecommand \@@endlink[0]{}%
\providecommand \url  [0]{\begingroup\@sanitize@url \@url }%
\providecommand \@url [1]{\endgroup\@href {#1}{\urlprefix }}%
\providecommand \urlprefix  [0]{URL }%
\providecommand \Eprint [0]{\href }%
\providecommand \doibase [0]{http://dx.doi.org/}%
\providecommand \selectlanguage [0]{\@gobble}%
\providecommand \bibinfo  [0]{\@secondoftwo}%
\providecommand \bibfield  [0]{\@secondoftwo}%
\providecommand \translation [1]{[#1]}%
\providecommand \BibitemOpen [0]{}%
\providecommand \bibitemStop [0]{}%
\providecommand \bibitemNoStop [0]{.\EOS\space}%
\providecommand \EOS [0]{\spacefactor3000\relax}%
\providecommand \BibitemShut  [1]{\csname bibitem#1\endcsname}%
\let\auto@bib@innerbib\@empty
\end{thebibliography}%


\begin{thebibliography}{42}%
\makeatletter
\providecommand \@ifxundefined [1]{%
 \@ifx{#1\undefined}
}%
\providecommand \@ifnum [1]{%
 \ifnum #1\expandafter \@firstoftwo
 \else \expandafter \@secondoftwo
 \fi
}%
\providecommand \@ifx [1]{%
 \ifx #1\expandafter \@firstoftwo
 \else \expandafter \@secondoftwo
 \fi
}%
\providecommand \natexlab [1]{#1}%
\providecommand \enquote  [1]{``#1''}%
\providecommand \bibnamefont  [1]{#1}%
\providecommand \bibfnamefont [1]{#1}%
\providecommand \citenamefont [1]{#1}%
\providecommand \href@noop [0]{\@secondoftwo}%
\providecommand \href [0]{\begingroup \@sanitize@url \@href}%
\providecommand \@href[1]{\@@startlink{#1}\@@href}%
\providecommand \@@href[1]{\endgroup#1\@@endlink}%
\providecommand \@sanitize@url [0]{\catcode `\\12\catcode `\$12\catcode
  `\&12\catcode `\#12\catcode `\^12\catcode `\_12\catcode `\%12\relax}%
\providecommand \@@startlink[1]{}%
\providecommand \@@endlink[0]{}%
\providecommand \url  [0]{\begingroup\@sanitize@url \@url }%
\providecommand \@url [1]{\endgroup\@href {#1}{\urlprefix }}%
\providecommand \urlprefix  [0]{URL }%
\providecommand \Eprint [0]{\href }%
\providecommand \doibase [0]{http://dx.doi.org/}%
\providecommand \selectlanguage [0]{\@gobble}%
\providecommand \bibinfo  [0]{\@secondoftwo}%
\providecommand \bibfield  [0]{\@secondoftwo}%
\providecommand \translation [1]{[#1]}%
\providecommand \BibitemOpen [0]{}%
\providecommand \bibitemStop [0]{}%
\providecommand \bibitemNoStop [0]{.\EOS\space}%
\providecommand \EOS [0]{\spacefactor3000\relax}%
\providecommand \BibitemShut  [1]{\csname bibitem#1\endcsname}%
\let\auto@bib@innerbib\@empty
\bibitem [{\citenamefont {Morko{\c{c}}}(2009)}]{Morkoc:2009_B}%
  \BibitemOpen
  \bibfield  {author} {\bibinfo {author} {\bibfnamefont {H.}~\bibnamefont
  {Morko{\c{c}}}},\ }\href@noop {} {\emph {\bibinfo {title} {Handbook of
  Nitride Semiconductors and Devices: GaN-based Optical and Electronic
  Devices}}},\ Vol.~\bibinfo {volume} {3}\ (\bibinfo  {publisher} {Wiley-VCH, Weinheim},\
  \bibinfo {year} {2009})\BibitemShut {NoStop}%
\bibitem [{\citenamefont {Palacios}\ \emph {et~al.}(2005)\citenamefont
  {Palacios}, \citenamefont {Chakraborty}, \citenamefont {Rajan}, \citenamefont
  {Poblenz}, \citenamefont {Keller}, \citenamefont {DenBaars}, \citenamefont
  {Speck},\ and\ \citenamefont {Mishra}}]{Palacios:2005_IEEE}%
  \BibitemOpen
  \bibfield  {author} {\bibinfo {author} {\bibfnamefont {T.}~\bibnamefont
  {Palacios}}, \bibinfo {author} {\bibfnamefont {A.}~\bibnamefont
  {Chakraborty}}, \bibinfo {author} {\bibfnamefont {S.}~\bibnamefont {Rajan}},
  \bibinfo {author} {\bibfnamefont {C.}~\bibnamefont {Poblenz}}, \bibinfo
  {author} {\bibfnamefont {S.}~\bibnamefont {Keller}}, \bibinfo {author}
  {\bibfnamefont {S.}~\bibnamefont {DenBaars}}, \bibinfo {author}
  {\bibfnamefont {J.}~\bibnamefont {Speck}}, \ and\ \bibinfo {author}
  {\bibfnamefont {U.}~\bibnamefont {Mishra}},\ }\href {\doibase
  10.1109/LED.2005.857701} {\bibfield  {journal} {\bibinfo  {journal} {IEEE
  Electron Device Lett.}\ }\textbf {\bibinfo {volume} {26}},\ \bibinfo {pages}
  {781} (\bibinfo {year} {2005})}\BibitemShut {NoStop}%
\bibitem [{\citenamefont {Kaun}\ \emph {et~al.}(2013)\citenamefont {Kaun},
  \citenamefont {Wong}, \citenamefont {Mishra},\ and\ \citenamefont
  {Speck}}]{Kaun:2013_SST}%
  \BibitemOpen
  \bibfield  {author} {\bibinfo {author} {\bibfnamefont {S.~W.}\ \bibnamefont
  {Kaun}}, \bibinfo {author} {\bibfnamefont {M.~H.}\ \bibnamefont {Wong}},
  \bibinfo {author} {\bibfnamefont {U.~K.}\ \bibnamefont {Mishra}}, \ and\
  \bibinfo {author} {\bibfnamefont {J.~S.}\ \bibnamefont {Speck}},\ }\href
  {\doibase 10.1088/0268-1242/28/7/074001} {\bibfield  {journal} {\bibinfo
  {journal} {Semicond. Sci. Technol.}\ }\textbf {\bibinfo {volume} {28}},\
  \bibinfo {pages} {074001} (\bibinfo {year} {2013})}\BibitemShut {NoStop}%
\bibitem [{\citenamefont {Bonanni}\ \emph {et~al.}(2008)\citenamefont
  {Bonanni}, \citenamefont {Navarro-Quezada}, \citenamefont {Li}, \citenamefont
  {Wegscheider}, \citenamefont {Mat\v{e}j}, \citenamefont {Hol\'y},
  \citenamefont {Lechner}, \citenamefont {Bauer}, \citenamefont {Rovezzi},
  \citenamefont {D'Acapito}, \citenamefont {Kiecana}, \citenamefont {Sawicki},\
  and\ \citenamefont {Dietl}}]{Bonanni:2008_PRL}%
  \BibitemOpen
  \bibfield  {author} {\bibinfo {author} {\bibfnamefont {A.}~\bibnamefont
  {Bonanni}}, \bibinfo {author} {\bibfnamefont {A.}~\bibnamefont
  {Navarro-Quezada}}, \bibinfo {author} {\bibfnamefont {T.}~\bibnamefont {Li}},
  \bibinfo {author} {\bibfnamefont {M.}~\bibnamefont {Wegscheider}}, \bibinfo
  {author} {\bibfnamefont {Z.}~\bibnamefont {Mat\v{e}j}}, \bibinfo {author}
  {\bibfnamefont {V.}~\bibnamefont {Hol\'y}}, \bibinfo {author} {\bibfnamefont
  {R.~T.}\ \bibnamefont {Lechner}}, \bibinfo {author} {\bibfnamefont
  {G.}~\bibnamefont {Bauer}}, \bibinfo {author} {\bibfnamefont
  {M.}~\bibnamefont {Rovezzi}}, \bibinfo {author} {\bibfnamefont
  {F.}~\bibnamefont {D'Acapito}}, \bibinfo {author} {\bibfnamefont
  {M.}~\bibnamefont {Kiecana}}, \bibinfo {author} {\bibfnamefont
  {M.}~\bibnamefont {Sawicki}}, \ and\ \bibinfo {author} {\bibfnamefont
  {T.}~\bibnamefont {Dietl}},\ }\href {\doibase 10.1103/PhysRevLett.101.135502}
  {\bibfield  {journal} {\bibinfo  {journal} {Phys. Rev. Lett.}\ }\textbf
  {\bibinfo {volume} {101}},\ \bibinfo {pages} {135502} (\bibinfo {year}
  {2008})}\BibitemShut {NoStop}%
\bibitem [{\citenamefont {Navarro-Quezada}\ \emph {et~al.}(2010)\citenamefont
  {Navarro-Quezada}, \citenamefont {Stefanowicz}, \citenamefont {Li},
  \citenamefont {Faina}, \citenamefont {Rovezzi}, \citenamefont {Lechner},
  \citenamefont {Devillers}, \citenamefont {d'Acapito}, \citenamefont {Bauer},
  \citenamefont {Sawicki}, \citenamefont {Dietl},\ and\ \citenamefont
  {Bonanni}}]{Navarro:2010_PRB}%
  \BibitemOpen
  \bibfield  {author} {\bibinfo {author} {\bibfnamefont {A.}~\bibnamefont
  {Navarro-Quezada}}, \bibinfo {author} {\bibfnamefont {W.}~\bibnamefont
  {Stefanowicz}}, \bibinfo {author} {\bibfnamefont {T.}~\bibnamefont {Li}},
  \bibinfo {author} {\bibfnamefont {B.}~\bibnamefont {Faina}}, \bibinfo
  {author} {\bibfnamefont {M.}~\bibnamefont {Rovezzi}}, \bibinfo {author}
  {\bibfnamefont {R.~T.}\ \bibnamefont {Lechner}}, \bibinfo {author}
  {\bibfnamefont {T.}~\bibnamefont {Devillers}}, \bibinfo {author}
  {\bibfnamefont {F.}~\bibnamefont {d'Acapito}}, \bibinfo {author}
  {\bibfnamefont {G.}~\bibnamefont {Bauer}}, \bibinfo {author} {\bibfnamefont
  {M.}~\bibnamefont {Sawicki}}, \bibinfo {author} {\bibfnamefont
  {T.}~\bibnamefont {Dietl}}, \ and\ \bibinfo {author} {\bibfnamefont
  {A.}~\bibnamefont {Bonanni}},\ }\href {\doibase 10.1103/PhysRevB.81.205206}
  {\bibfield  {journal} {\bibinfo  {journal} {Phys. Rev. B}\ }\textbf {\bibinfo
  {volume} {81}},\ \bibinfo {pages} {205206} (\bibinfo {year}
  {2010})}\BibitemShut {NoStop}%
\bibitem [{\citenamefont {Sheu}\ \emph {et~al.}(2013)\citenamefont {Sheu},
  \citenamefont {Huang}, \citenamefont {Lee}, \citenamefont {Lee},
  \citenamefont {Yeh}, \citenamefont {Chen},\ and\ \citenamefont
  {Lai}}]{Sheu:2013_APL}%
  \BibitemOpen
  \bibfield  {author} {\bibinfo {author} {\bibfnamefont {J.-K.}\ \bibnamefont
  {Sheu}}, \bibinfo {author} {\bibfnamefont {F.-W.}\ \bibnamefont {Huang}},
  \bibinfo {author} {\bibfnamefont {C.-H.}\ \bibnamefont {Lee}}, \bibinfo
  {author} {\bibfnamefont {M.-L.}\ \bibnamefont {Lee}}, \bibinfo {author}
  {\bibfnamefont {Y.-H.}\ \bibnamefont {Yeh}}, \bibinfo {author} {\bibfnamefont
  {P.-C.}\ \bibnamefont {Chen}}, \ and\ \bibinfo {author} {\bibfnamefont
  {W.-C.}\ \bibnamefont {Lai}},\ }\href@noop {} {\bibfield  {journal} {\bibinfo
   {journal} {Appl. Phys. Lett.}\ }\textbf {\bibinfo {volume} {103}},\ \bibinfo
  {pages} {063906} (\bibinfo {year} {2013})}\BibitemShut {NoStop}%
\bibitem [{\citenamefont {Devillers}\ \emph {et~al.}(2012)\citenamefont
  {Devillers}, \citenamefont {Rovezzi}, \citenamefont {Szwacki}, \citenamefont
  {Dobkowska}, \citenamefont {Stefanowicz}, \citenamefont {Sztenkiel},
  \citenamefont {Grois}, \citenamefont {Suffczy{\'n}ski}, \citenamefont
  {Navarro-Quezada}, \citenamefont {Faina}, \citenamefont {Glatzel},
  \citenamefont {d'Acapito}, \citenamefont {Jakie{\l}a}, \citenamefont
  {Sawicki}, \citenamefont {Majewski}, \citenamefont {Dietl},\ and\
  \citenamefont {Bonanni}}]{Devillers:2012_SR}%
  \BibitemOpen
  \bibfield  {author} {\bibinfo {author} {\bibfnamefont {T.}~\bibnamefont
  {Devillers}}, \bibinfo {author} {\bibfnamefont {M.}~\bibnamefont {Rovezzi}},
  \bibinfo {author} {\bibfnamefont {N.~G.}\ \bibnamefont {Szwacki}}, \bibinfo
  {author} {\bibfnamefont {S.}~\bibnamefont {Dobkowska}}, \bibinfo {author}
  {\bibfnamefont {W.}~\bibnamefont {Stefanowicz}}, \bibinfo {author}
  {\bibfnamefont {D.}~\bibnamefont {Sztenkiel}}, \bibinfo {author}
  {\bibfnamefont {A.}~\bibnamefont {Grois}}, \bibinfo {author} {\bibfnamefont
  {J.}~\bibnamefont {Suffczy{\'n}ski}}, \bibinfo {author} {\bibfnamefont
  {A.}~\bibnamefont {Navarro-Quezada}}, \bibinfo {author} {\bibfnamefont
  {T.}~\bibnamefont {Faina}, \bibfnamefont {B.~Li}}, \bibinfo {author}
  {\bibfnamefont {P.}~\bibnamefont {Glatzel}}, \bibinfo {author} {\bibfnamefont
  {F.}~\bibnamefont {d'Acapito}}, \bibinfo {author} {\bibfnamefont
  {R.}~\bibnamefont {Jakie{\l}a}}, \bibinfo {author} {\bibfnamefont
  {M.}~\bibnamefont {Sawicki}}, \bibinfo {author} {\bibfnamefont {J.~A.}\
  \bibnamefont {Majewski}}, \bibinfo {author} {\bibfnamefont {T.}~\bibnamefont
  {Dietl}}, \ and\ \bibinfo {author} {\bibfnamefont {A.}~\bibnamefont
  {Bonanni}},\ }\href@noop {} {\bibfield  {journal} {\bibinfo  {journal} {Sci.
  Rep.}\ }\textbf {\bibinfo {volume} {2}} (\bibinfo {year} {2012})}\BibitemShut
  {NoStop}%
\bibitem [{\citenamefont {Beschoten}\ \emph {et~al.}(2001)\citenamefont
  {Beschoten}, \citenamefont {Johnston-Halperin}, \citenamefont {Young},
  \citenamefont {Poggio}, \citenamefont {Grimaldi}, \citenamefont {Keller},
  \citenamefont {DenBaars}, \citenamefont {Mishra}, \citenamefont {Hu},\ and\
  \citenamefont {Awschalom}}]{Beschoten:2001_PRB}%
  \BibitemOpen
  \bibfield  {author} {\bibinfo {author} {\bibfnamefont {B.}~\bibnamefont
  {Beschoten}}, \bibinfo {author} {\bibfnamefont {E.}~\bibnamefont
  {Johnston-Halperin}}, \bibinfo {author} {\bibfnamefont {D.~K.}\ \bibnamefont
  {Young}}, \bibinfo {author} {\bibfnamefont {M.}~\bibnamefont {Poggio}},
  \bibinfo {author} {\bibfnamefont {J.~E.}\ \bibnamefont {Grimaldi}}, \bibinfo
  {author} {\bibfnamefont {S.}~\bibnamefont {Keller}}, \bibinfo {author}
  {\bibfnamefont {S.~P.}\ \bibnamefont {DenBaars}}, \bibinfo {author}
  {\bibfnamefont {U.~K.}\ \bibnamefont {Mishra}}, \bibinfo {author}
  {\bibfnamefont {E.~L.}\ \bibnamefont {Hu}}, \ and\ \bibinfo {author}
  {\bibfnamefont {D.~D.}\ \bibnamefont {Awschalom}},\ }\href {\doibase
  10.1103/PhysRevB.63.121202} {\bibfield  {journal} {\bibinfo  {journal} {Phys.
  Rev. B}\ }\textbf {\bibinfo {volume} {63}},\ \bibinfo {pages} {121202}
  (\bibinfo {year} {2001})}\BibitemShut {NoStop}%
\bibitem [{\citenamefont {Krishnamurthy}\ \emph {et~al.}(2003)\citenamefont
  {Krishnamurthy}, \citenamefont {van Schilfgaarde},\ and\ \citenamefont
  {Newman}}]{Krishnamurthy:2003_APL}%
  \BibitemOpen
  \bibfield  {author} {\bibinfo {author} {\bibfnamefont {S.}~\bibnamefont
  {Krishnamurthy}}, \bibinfo {author} {\bibfnamefont {M.}~\bibnamefont {van
  Schilfgaarde}}, \ and\ \bibinfo {author} {\bibfnamefont {N.}~\bibnamefont
  {Newman}},\ }\href {\doibase http://dx.doi.org/10.1063/1.1606873} {\bibfield
  {journal} {\bibinfo  {journal} {Appl. Phys. Lett.}\ }\textbf {\bibinfo
  {volume} {83}},\ \bibinfo {pages} {1761} (\bibinfo {year}
  {2003})}\BibitemShut {NoStop}%
\bibitem [{\citenamefont {Miao}\ \emph {et~al.}(2012)\citenamefont {Miao},
  \citenamefont {Yan}, \citenamefont {Van~de Walle}, \citenamefont {Lou},
  \citenamefont {Li},\ and\ \citenamefont {Chang}}]{Miao:2012_PRL}%
  \BibitemOpen
  \bibfield  {author} {\bibinfo {author} {\bibfnamefont {M.~S.}\ \bibnamefont
  {Miao}}, \bibinfo {author} {\bibfnamefont {Q.}~\bibnamefont {Yan}}, \bibinfo
  {author} {\bibfnamefont {C.~G.}\ \bibnamefont {Van~de Walle}}, \bibinfo
  {author} {\bibfnamefont {W.~K.}\ \bibnamefont {Lou}}, \bibinfo {author}
  {\bibfnamefont {L.~L.}\ \bibnamefont {Li}}, \ and\ \bibinfo {author}
  {\bibfnamefont {K.}~\bibnamefont {Chang}},\ }\href {\doibase
  10.1103/PhysRevLett.109.186803} {\bibfield  {journal} {\bibinfo  {journal}
  {Phys. Rev. Lett.}\ }\textbf {\bibinfo {volume} {109}},\ \bibinfo {pages}
  {186803} (\bibinfo {year} {2012})}\BibitemShut {NoStop}%
\bibitem [{\citenamefont {Bonanni}\ \emph {et~al.}(2011)\citenamefont
  {Bonanni}, \citenamefont {Sawicki}, \citenamefont {Devillers}, \citenamefont
  {Stefanowicz}, \citenamefont {Faina}, \citenamefont {Li}, \citenamefont
  {Winkler}, \citenamefont {Sztenkiel}, \citenamefont {Navarro-Quezada},
  \citenamefont {Rovezzi}, \citenamefont {Jakie{\l}a}, \citenamefont {Grois},
  \citenamefont {Wegscheider}, \citenamefont {Jantsch}, \citenamefont
  {Suffczy\'nski}, \citenamefont {D'Acapito}, \citenamefont {Meingast},
  \citenamefont {Kothleitner},\ and\ \citenamefont {Dietl}}]{Bonanni:2011_PRB}%
  \BibitemOpen
  \bibfield  {author} {\bibinfo {author} {\bibfnamefont {A.}~\bibnamefont
  {Bonanni}}, \bibinfo {author} {\bibfnamefont {M.}~\bibnamefont {Sawicki}},
  \bibinfo {author} {\bibfnamefont {T.}~\bibnamefont {Devillers}}, \bibinfo
  {author} {\bibfnamefont {W.}~\bibnamefont {Stefanowicz}}, \bibinfo {author}
  {\bibfnamefont {B.}~\bibnamefont {Faina}}, \bibinfo {author} {\bibfnamefont
  {T.}~\bibnamefont {Li}}, \bibinfo {author} {\bibfnamefont {T.~E.}\
  \bibnamefont {Winkler}}, \bibinfo {author} {\bibfnamefont {D.}~\bibnamefont
  {Sztenkiel}}, \bibinfo {author} {\bibfnamefont {A.}~\bibnamefont
  {Navarro-Quezada}}, \bibinfo {author} {\bibfnamefont {M.}~\bibnamefont
  {Rovezzi}}, \bibinfo {author} {\bibfnamefont {R.}~\bibnamefont {Jakie{\l}a}},
  \bibinfo {author} {\bibfnamefont {A.}~\bibnamefont {Grois}}, \bibinfo
  {author} {\bibfnamefont {M.}~\bibnamefont {Wegscheider}}, \bibinfo {author}
  {\bibfnamefont {W.}~\bibnamefont {Jantsch}}, \bibinfo {author} {\bibfnamefont
  {J.}~\bibnamefont {Suffczy\'nski}}, \bibinfo {author} {\bibfnamefont
  {F.}~\bibnamefont {D'Acapito}}, \bibinfo {author} {\bibfnamefont
  {A.}~\bibnamefont {Meingast}}, \bibinfo {author} {\bibfnamefont
  {G.}~\bibnamefont {Kothleitner}}, \ and\ \bibinfo {author} {\bibfnamefont
  {T.}~\bibnamefont {Dietl}},\ }\href {\doibase 10.1103/PhysRevB.84.035206}
  {\bibfield  {journal} {\bibinfo  {journal} {Phys. Rev. B}\ }\textbf {\bibinfo
  {volume} {84}},\ \bibinfo {pages} {035206} (\bibinfo {year}
  {2011})}\BibitemShut {NoStop}%
\bibitem [{\citenamefont {Sawicki}\ \emph {et~al.}(2012)\citenamefont
  {Sawicki}, \citenamefont {Devillers}, \citenamefont {Ga\l{\c{e}}ski},
  \citenamefont {Simserides}, \citenamefont {Dobkowska}, \citenamefont {Faina},
  \citenamefont {Grois}, \citenamefont {Navarro-Quezada}, \citenamefont
  {Trohidou}, \citenamefont {Majewski}, \citenamefont {Dietl},\ and\
  \citenamefont {Bonanni}}]{Sawicki:2012_PRB}%
  \BibitemOpen
  \bibfield  {author} {\bibinfo {author} {\bibfnamefont {M.}~\bibnamefont
  {Sawicki}}, \bibinfo {author} {\bibfnamefont {T.}~\bibnamefont {Devillers}},
  \bibinfo {author} {\bibfnamefont {S.}~\bibnamefont {Ga\l{\c{e}}ski}},
  \bibinfo {author} {\bibfnamefont {C.}~\bibnamefont {Simserides}}, \bibinfo
  {author} {\bibfnamefont {S.}~\bibnamefont {Dobkowska}}, \bibinfo {author}
  {\bibfnamefont {B.}~\bibnamefont {Faina}}, \bibinfo {author} {\bibfnamefont
  {A.}~\bibnamefont {Grois}}, \bibinfo {author} {\bibfnamefont
  {A.}~\bibnamefont {Navarro-Quezada}}, \bibinfo {author} {\bibfnamefont
  {K.~N.}\ \bibnamefont {Trohidou}}, \bibinfo {author} {\bibfnamefont {J.~A.}\
  \bibnamefont {Majewski}}, \bibinfo {author} {\bibfnamefont {T.}~\bibnamefont
  {Dietl}}, \ and\ \bibinfo {author} {\bibfnamefont {A.}~\bibnamefont
  {Bonanni}},\ }\href {\doibase 10.1103/PhysRevB.85.205204} {\bibfield
  {journal} {\bibinfo  {journal} {Phys. Rev. B}\ }\textbf {\bibinfo {volume}
  {85}},\ \bibinfo {pages} {205204} (\bibinfo {year} {2012})}\BibitemShut
  {NoStop}%
\bibitem [{\citenamefont {Stefanowicz}\ \emph {et~al.}(2013)\citenamefont
  {Stefanowicz}, \citenamefont {Kunert}, \citenamefont {Simserides},
  \citenamefont {Majewski}, \citenamefont {Stefanowicz}, \citenamefont {Kruse},
  \citenamefont {Figge}, \citenamefont {Li}, \citenamefont {Jakie{\l}a},
  \citenamefont {Trohidou}, \citenamefont {Bonanni}, \citenamefont {Hommel},
  \citenamefont {Sawicki},\ and\ \citenamefont {Dietl}}]{Stefanowicz:2013_PRB}%
  \BibitemOpen
  \bibfield  {author} {\bibinfo {author} {\bibfnamefont {S.}~\bibnamefont
  {Stefanowicz}}, \bibinfo {author} {\bibfnamefont {G.}~\bibnamefont {Kunert}},
  \bibinfo {author} {\bibfnamefont {C.}~\bibnamefont {Simserides}}, \bibinfo
  {author} {\bibfnamefont {J.~A.}\ \bibnamefont {Majewski}}, \bibinfo {author}
  {\bibfnamefont {W.}~\bibnamefont {Stefanowicz}}, \bibinfo {author}
  {\bibfnamefont {C.}~\bibnamefont {Kruse}}, \bibinfo {author} {\bibfnamefont
  {S.}~\bibnamefont {Figge}}, \bibinfo {author} {\bibfnamefont
  {T.}~\bibnamefont {Li}}, \bibinfo {author} {\bibfnamefont {R.}~\bibnamefont
  {Jakie{\l}a}}, \bibinfo {author} {\bibfnamefont {K.~N.}\ \bibnamefont
  {Trohidou}}, \bibinfo {author} {\bibfnamefont {A.}~\bibnamefont {Bonanni}},
  \bibinfo {author} {\bibfnamefont {D.}~\bibnamefont {Hommel}}, \bibinfo
  {author} {\bibfnamefont {M.}~\bibnamefont {Sawicki}}, \ and\ \bibinfo
  {author} {\bibfnamefont {T.}~\bibnamefont {Dietl}},\ }\href {\doibase
  10.1103/PhysRevB.88.081201} {\bibfield  {journal} {\bibinfo  {journal} {Phys.
  Rev. B}\ }\textbf {\bibinfo {volume} {88}},\ \bibinfo {pages} {081201}
  (\bibinfo {year} {2013})}\BibitemShut {NoStop}%
\bibitem [{\citenamefont {Sau}\ \emph {et~al.}(2010)\citenamefont {Sau},
  \citenamefont {Lutchyn}, \citenamefont {Tewari},\ and\ \citenamefont
  {Das~Sarma}}]{Sau:2010_PRL}%
  \BibitemOpen
  \bibfield  {author} {\bibinfo {author} {\bibfnamefont {J.~D.}\ \bibnamefont
  {Sau}}, \bibinfo {author} {\bibfnamefont {R.~M.}\ \bibnamefont {Lutchyn}},
  \bibinfo {author} {\bibfnamefont {S.}~\bibnamefont {Tewari}}, \ and\ \bibinfo
  {author} {\bibfnamefont {S.}~\bibnamefont {Das~Sarma}},\ }\href {\doibase
  10.1103/PhysRevLett.104.040502} {\bibfield  {journal} {\bibinfo  {journal}
  {Phys. Rev. Lett.}\ }\textbf {\bibinfo {volume} {104}},\ \bibinfo {pages}
  {040502} (\bibinfo {year} {2010})}\BibitemShut {NoStop}%
\bibitem [{\citenamefont {Hikami}\ \emph {et~al.}(1980)\citenamefont {Hikami},
  \citenamefont {Larkin},\ and\ \citenamefont {Nagaoka}}]{Hikami:1980_PTP}%
  \BibitemOpen
  \bibfield  {author} {\bibinfo {author} {\bibfnamefont {S.}~\bibnamefont
  {Hikami}}, \bibinfo {author} {\bibfnamefont {A.~I.}\ \bibnamefont {Larkin}},
  \ and\ \bibinfo {author} {\bibfnamefont {Y.}~\bibnamefont {Nagaoka}},\ }\href
  {\doibase 10.1143/PTP.63.707} {\bibfield  {journal} {\bibinfo  {journal}
  {Prog. Theo. Phys.}\ }\textbf {\bibinfo {volume} {63}},\ \bibinfo {pages}
  {707} (\bibinfo {year} {1980})}\BibitemShut {NoStop}%
\bibitem [{\citenamefont {Sawicki}\ \emph {et~al.}(1986)\citenamefont
  {Sawicki}, \citenamefont {Dietl}, \citenamefont {Kossut}, \citenamefont
  {Igalson}, \citenamefont {Wojtowicz},\ and\ \citenamefont
  {Plesiewicz}}]{Sawicki:1986_PRL}%
  \BibitemOpen
  \bibfield  {author} {\bibinfo {author} {\bibfnamefont {M.}~\bibnamefont
  {Sawicki}}, \bibinfo {author} {\bibfnamefont {T.}~\bibnamefont {Dietl}},
  \bibinfo {author} {\bibfnamefont {J.}~\bibnamefont {Kossut}}, \bibinfo
  {author} {\bibfnamefont {J.}~\bibnamefont {Igalson}}, \bibinfo {author}
  {\bibfnamefont {T.}~\bibnamefont {Wojtowicz}}, \ and\ \bibinfo {author}
  {\bibfnamefont {W.}~\bibnamefont {Plesiewicz}},\ }\href {\doibase
  10.1103/PhysRevLett.56.508} {\bibfield  {journal} {\bibinfo  {journal} {Phys.
  Rev. Lett.}\ }\textbf {\bibinfo {volume} {56}},\ \bibinfo {pages} {508}
  (\bibinfo {year} {1986})}\BibitemShut {NoStop}%
\bibitem [{\citenamefont {Rashba}(1960)}]{Rashba:1960_SPSS}%
  \BibitemOpen
  \bibfield  {author} {\bibinfo {author} {\bibfnamefont {E.~I.}\ \bibnamefont
  {Rashba}},\ }\href@noop {} {\bibfield  {journal} {\bibinfo  {journal} {Sov.
  Phys. Solid State}\ }\textbf {\bibinfo {volume} {2}},\ \bibinfo {pages}
  {1109} (\bibinfo {year} {1960})}\BibitemShut {NoStop}%
\bibitem [{\citenamefont {Bychkov}\ and\ \citenamefont
  {Rashba}(1984)}]{Bychkov:1984_JPC}%
  \BibitemOpen
  \bibfield  {author} {\bibinfo {author} {\bibfnamefont {Y.~A.}\ \bibnamefont
  {Bychkov}}\ and\ \bibinfo {author} {\bibfnamefont {E.~I.}\ \bibnamefont
  {Rashba}},\ }\href {\doibase 10.1088/0022-3719/17/33/015} {\bibfield
  {journal} {\bibinfo  {journal} {J. Phys. C}\ }\textbf {\bibinfo {volume}
  {17}},\ \bibinfo {pages} {6039} (\bibinfo {year} {1984})}\BibitemShut
  {NoStop}%
\bibitem [{\citenamefont {Wolos}\ \emph {et~al.}(2011)\citenamefont {Wolos},
  \citenamefont {Wilamowski}, \citenamefont {Skierbiszewski}, \citenamefont
  {Drabinska}, \citenamefont {Lucznik}, \citenamefont {Grzegory},\ and\
  \citenamefont {Porowski}}]{Wolos:2011_PB}%
  \BibitemOpen
  \bibfield  {author} {\bibinfo {author} {\bibfnamefont {A.}~\bibnamefont
  {Wolos}}, \bibinfo {author} {\bibfnamefont {Z.}~\bibnamefont {Wilamowski}},
  \bibinfo {author} {\bibfnamefont {C.}~\bibnamefont {Skierbiszewski}},
  \bibinfo {author} {\bibfnamefont {A.}~\bibnamefont {Drabinska}}, \bibinfo
  {author} {\bibfnamefont {B.}~\bibnamefont {Lucznik}}, \bibinfo {author}
  {\bibfnamefont {I.}~\bibnamefont {Grzegory}}, \ and\ \bibinfo {author}
  {\bibfnamefont {S.}~\bibnamefont {Porowski}},\ }\href {\doibase
  http://dx.doi.org/10.1016/j.physb.2011.03.060} {\bibfield  {journal}
  {\bibinfo  {journal} {Physica B}\ }\textbf {\bibinfo {volume} {406}},\
  \bibinfo {pages} {2548 } (\bibinfo {year} {2011})}\BibitemShut {NoStop}%
\bibitem [{\citenamefont {Schmult}\ \emph {et~al.}(2006)\citenamefont
  {Schmult}, \citenamefont {Manfra}, \citenamefont {Punnoose}, \citenamefont
  {Sergent}, \citenamefont {Baldwin},\ and\ \citenamefont
  {Molnar}}]{Schmult:2006_PRB}%
  \BibitemOpen
  \bibfield  {author} {\bibinfo {author} {\bibfnamefont {S.}~\bibnamefont
  {Schmult}}, \bibinfo {author} {\bibfnamefont {M.~J.}\ \bibnamefont {Manfra}},
  \bibinfo {author} {\bibfnamefont {A.}~\bibnamefont {Punnoose}}, \bibinfo
  {author} {\bibfnamefont {A.~M.}\ \bibnamefont {Sergent}}, \bibinfo {author}
  {\bibfnamefont {K.~W.}\ \bibnamefont {Baldwin}}, \ and\ \bibinfo {author}
  {\bibfnamefont {R.~J.}\ \bibnamefont {Molnar}},\ }\href {\doibase
  10.1103/PhysRevB.74.033302} {\bibfield  {journal} {\bibinfo  {journal} {Phys.
  Rev. B}\ }\textbf {\bibinfo {volume} {74}},\ \bibinfo {pages} {033302}
  (\bibinfo {year} {2006})}\BibitemShut {NoStop}%
\bibitem [{\citenamefont {Kurdak}\ \emph {et~al.}(2006)\citenamefont {Kurdak},
  \citenamefont {Biyikli}, \citenamefont {\"Ozg\"ur}, \citenamefont
  {Morko\c{c}},\ and\ \citenamefont {Litvinov}}]{Kurdak:2006_PRB}%
  \BibitemOpen
  \bibfield  {author} {\bibinfo {author} {\bibfnamefont {C.}~\bibnamefont
  {Kurdak}}, \bibinfo {author} {\bibfnamefont {N.}~\bibnamefont {Biyikli}},
  \bibinfo {author} {\bibfnamefont {U.}~\bibnamefont {\"Ozg\"ur}}, \bibinfo
  {author} {\bibfnamefont {H.}~\bibnamefont {Morko\c{c}}}, \ and\ \bibinfo
  {author} {\bibfnamefont {V.~I.}\ \bibnamefont {Litvinov}},\ }\href {\doibase
  10.1103/PhysRevB.74.113308} {\bibfield  {journal} {\bibinfo  {journal} {Phys.
  Rev. B}\ }\textbf {\bibinfo {volume} {74}},\ \bibinfo {pages} {113308}
  (\bibinfo {year} {2006})}\BibitemShut {NoStop}%
\bibitem [{\citenamefont {Thillosen}\ \emph
  {et~al.}(2006{\natexlab{a}})\citenamefont {Thillosen}, \citenamefont
  {Sch\"apers}, \citenamefont {Kaluza}, \citenamefont {Hardtdegen},\ and\
  \citenamefont {Guzenko}}]{Thillosen:2006_APL}%
  \BibitemOpen
  \bibfield  {author} {\bibinfo {author} {\bibfnamefont {N.}~\bibnamefont
  {Thillosen}}, \bibinfo {author} {\bibfnamefont {T.}~\bibnamefont
  {Sch\"apers}}, \bibinfo {author} {\bibfnamefont {N.}~\bibnamefont {Kaluza}},
  \bibinfo {author} {\bibfnamefont {H.}~\bibnamefont {Hardtdegen}}, \ and\
  \bibinfo {author} {\bibfnamefont {V.~A.}\ \bibnamefont {Guzenko}},\ }\href
  {\doibase http://dx.doi.org/10.1063/1.2162871} {\bibfield  {journal}
  {\bibinfo  {journal} {Appl. Phys. Lett.}\ }\textbf {\bibinfo {volume} {88}},\
  \bibinfo {eid} {022111} (\bibinfo {year} {2006}{\natexlab{a}})}\BibitemShut
  {NoStop}%
\bibitem [{\citenamefont {Thillosen}\ \emph
  {et~al.}(2006{\natexlab{b}})\citenamefont {Thillosen}, \citenamefont
  {Caba\~nas}, \citenamefont {Kaluza}, \citenamefont {Guzenko}, \citenamefont
  {Hardtdegen},\ and\ \citenamefont {Sch\"apers}}]{Thillosen:2006_PRB}%
  \BibitemOpen
  \bibfield  {author} {\bibinfo {author} {\bibfnamefont {N.}~\bibnamefont
  {Thillosen}}, \bibinfo {author} {\bibfnamefont {S.}~\bibnamefont
  {Caba\~nas}}, \bibinfo {author} {\bibfnamefont {N.}~\bibnamefont {Kaluza}},
  \bibinfo {author} {\bibfnamefont {V.~A.}\ \bibnamefont {Guzenko}}, \bibinfo
  {author} {\bibfnamefont {H.}~\bibnamefont {Hardtdegen}}, \ and\ \bibinfo
  {author} {\bibfnamefont {T.}~\bibnamefont {Sch\"apers}},\ }\href {\doibase
  10.1103/PhysRevB.73.241311} {\bibfield  {journal} {\bibinfo  {journal} {Phys.
  Rev. B}\ }\textbf {\bibinfo {volume} {73}},\ \bibinfo {pages} {241311}
  (\bibinfo {year} {2006}{\natexlab{b}})}\BibitemShut {NoStop}%
\bibitem [{\citenamefont {Zhou}\ \emph {et~al.}(2008)\citenamefont {Zhou},
  \citenamefont {Lin}, \citenamefont {Shang}, \citenamefont {Sun},
  \citenamefont {Gao}, \citenamefont {Zhou}, \citenamefont {Yu}, \citenamefont
  {Tang}, \citenamefont {Han}, \citenamefont {Shen}, \citenamefont {Guo},
  \citenamefont {Gui},\ and\ \citenamefont {Chu}}]{Zhou:2008_JAP}%
  \BibitemOpen
  \bibfield  {author} {\bibinfo {author} {\bibfnamefont {W.~Z.}\ \bibnamefont
  {Zhou}}, \bibinfo {author} {\bibfnamefont {T.}~\bibnamefont {Lin}}, \bibinfo
  {author} {\bibfnamefont {L.~Y.}\ \bibnamefont {Shang}}, \bibinfo {author}
  {\bibfnamefont {L.}~\bibnamefont {Sun}}, \bibinfo {author} {\bibfnamefont
  {K.~H.}\ \bibnamefont {Gao}}, \bibinfo {author} {\bibfnamefont {Y.~M.}\
  \bibnamefont {Zhou}}, \bibinfo {author} {\bibfnamefont {G.}~\bibnamefont
  {Yu}}, \bibinfo {author} {\bibfnamefont {N.}~\bibnamefont {Tang}}, \bibinfo
  {author} {\bibfnamefont {K.}~\bibnamefont {Han}}, \bibinfo {author}
  {\bibfnamefont {B.}~\bibnamefont {Shen}}, \bibinfo {author} {\bibfnamefont
  {S.~L.}\ \bibnamefont {Guo}}, \bibinfo {author} {\bibfnamefont {Y.~S.}\
  \bibnamefont {Gui}}, \ and\ \bibinfo {author} {\bibfnamefont {J.~H.}\
  \bibnamefont {Chu}},\ }\href {\doibase http://dx.doi.org/10.1063/1.2974091}
  {\bibfield  {journal} {\bibinfo  {journal} {J. Appl. Phys.}\ }\textbf
  {\bibinfo {volume} {104}},\ \bibinfo {eid} {053703} (\bibinfo {year}
  {2008})}\BibitemShut {NoStop}%
\bibitem [{\citenamefont {Cheng}\ \emph {et~al.}(2008)\citenamefont {Cheng},
  \citenamefont {Biyikli}, \citenamefont {\"Ozg\"ur}, \citenamefont {Kurdak},
  \citenamefont {Morko{\c{c}}},\ and\ \citenamefont
  {Litvinov}}]{Cheng:2008_PE}%
  \BibitemOpen
  \bibfield  {author} {\bibinfo {author} {\bibfnamefont {H.}~\bibnamefont
  {Cheng}}, \bibinfo {author} {\bibfnamefont {N.}~\bibnamefont {Biyikli}},
  \bibinfo {author} {\bibnamefont {\"Ozg\"ur}}, \bibinfo {author}
  {\bibfnamefont {{\c{C}}.}~\bibnamefont {Kurdak}}, \bibinfo {author}
  {\bibfnamefont {H.}~\bibnamefont {Morko{\c{c}}}}, \ and\ \bibinfo {author}
  {\bibfnamefont {V.~I.}\ \bibnamefont {Litvinov}},\ }\href {\doibase
  10.1016/j.physe.2007.09.184} {\bibfield  {journal} {\bibinfo  {journal}
  {Physica E}\ }\textbf {\bibinfo {volume} {40}},\ \bibinfo {pages} {1586}
  (\bibinfo {year} {2008})}\BibitemShut {NoStop}%
\bibitem [{\citenamefont {Belyaev}\ \emph {et~al.}(2008)\citenamefont
  {Belyaev}, \citenamefont {Raicheva}, \citenamefont {Kurakin}, \citenamefont
  {Klein},\ and\ \citenamefont {Vitusevich}}]{Belyaev:2008_PRB}%
  \BibitemOpen
  \bibfield  {author} {\bibinfo {author} {\bibfnamefont {A.~E.}\ \bibnamefont
  {Belyaev}}, \bibinfo {author} {\bibfnamefont {V.~G.}\ \bibnamefont
  {Raicheva}}, \bibinfo {author} {\bibfnamefont {A.~M.}\ \bibnamefont
  {Kurakin}}, \bibinfo {author} {\bibfnamefont {N.}~\bibnamefont {Klein}}, \
  and\ \bibinfo {author} {\bibfnamefont {S.~A.}\ \bibnamefont {Vitusevich}},\
  }\href {\doibase 10.1103/PhysRevB.77.035311} {\bibfield  {journal} {\bibinfo
  {journal} {Phys. Rev. B}\ }\textbf {\bibinfo {volume} {77}},\ \bibinfo
  {pages} {035311} (\bibinfo {year} {2008})}\BibitemShut {NoStop}%
\bibitem [{\citenamefont {Lew Yan~Voon}\ \emph {et~al.}(1996)\citenamefont {Lew
  Yan~Voon}, \citenamefont {Willatzen}, \citenamefont {Cardona},\ and\
  \citenamefont {Christensen}}]{Voon:1996_PRB}%
  \BibitemOpen
  \bibfield  {author} {\bibinfo {author} {\bibfnamefont {L.~C.}\ \bibnamefont
  {Lew Yan~Voon}}, \bibinfo {author} {\bibfnamefont {M.}~\bibnamefont
  {Willatzen}}, \bibinfo {author} {\bibfnamefont {M.}~\bibnamefont {Cardona}},
  \ and\ \bibinfo {author} {\bibfnamefont {N.~E.}\ \bibnamefont
  {Christensen}},\ }\href {\doibase 10.1103/PhysRevB.53.10703} {\bibfield
  {journal} {\bibinfo  {journal} {Phys. Rev. B}\ }\textbf {\bibinfo {volume}
  {53}},\ \bibinfo {pages} {10703} (\bibinfo {year} {1996})}\BibitemShut
  {NoStop}%
\bibitem [{\citenamefont {Majewski}\ and\ \citenamefont
  {Vogl}(2005)}]{Majewski:2005_P}%
  \BibitemOpen
  \bibfield  {author} {\bibinfo {author} {\bibfnamefont {J.~A.}\ \bibnamefont
  {Majewski}}\ and\ \bibinfo {author} {\bibfnamefont {P.}~\bibnamefont
  {Vogl}},\ }\href {\doibase http://dx.doi.org/10.1063/1.1994639} {\bibfield
  {journal} {\bibinfo  {journal} {AIP Conf. Proc.}\ }\textbf {\bibinfo {volume}
  {772}},\ \bibinfo {pages} {1403} (\bibinfo {year} {2005})}\BibitemShut
  {NoStop}%
\bibitem [{\citenamefont {Andrearczyk}\ \emph {et~al.}(2005)\citenamefont
  {Andrearczyk}, \citenamefont {Jaroszy\ifmmode~\acute{n}\else \'{n}\fi{}ski},
  \citenamefont {Grabecki}, \citenamefont {Dietl}, \citenamefont {Fukumura},\
  and\ \citenamefont {Kawasaki}}]{Andrearczyk:2005_PRB}%
  \BibitemOpen
  \bibfield  {author} {\bibinfo {author} {\bibfnamefont {T.}~\bibnamefont
  {Andrearczyk}}, \bibinfo {author} {\bibfnamefont {J.}~\bibnamefont
  {Jaroszy\ifmmode~\acute{n}\else \'{n}\fi{}ski}}, \bibinfo {author}
  {\bibfnamefont {G.}~\bibnamefont {Grabecki}}, \bibinfo {author}
  {\bibfnamefont {T.}~\bibnamefont {Dietl}}, \bibinfo {author} {\bibfnamefont
  {T.}~\bibnamefont {Fukumura}}, \ and\ \bibinfo {author} {\bibfnamefont
  {M.}~\bibnamefont {Kawasaki}},\ }\href {\doibase 10.1103/PhysRevB.72.121309}
  {\bibfield  {journal} {\bibinfo  {journal} {Phys. Rev. B}\ }\textbf {\bibinfo
  {volume} {72}},\ \bibinfo {pages} {121309} (\bibinfo {year}
  {2005})}\BibitemShut {NoStop}%
\bibitem [{\citenamefont {Romestain}\ \emph {et~al.}(1977)\citenamefont
  {Romestain}, \citenamefont {Geschwind},\ and\ \citenamefont
  {Devlin}}]{Romestain:1977_PRL}%
  \BibitemOpen
  \bibfield  {author} {\bibinfo {author} {\bibfnamefont {R.}~\bibnamefont
  {Romestain}}, \bibinfo {author} {\bibfnamefont {S.}~\bibnamefont
  {Geschwind}}, \ and\ \bibinfo {author} {\bibfnamefont {G.~E.}\ \bibnamefont
  {Devlin}},\ }\href {\doibase 10.1103/PhysRevLett.39.1583} {\bibfield
  {journal} {\bibinfo  {journal} {Phys. Rev. Lett.}\ }\textbf {\bibinfo
  {volume} {39}},\ \bibinfo {pages} {1583} (\bibinfo {year}
  {1977})}\BibitemShut {NoStop}%
\bibitem [{\citenamefont {Dobrowolska}\ \emph {et~al.}(1984)\citenamefont
  {Dobrowolska}, \citenamefont {Witowski}, \citenamefont {Furdyna},
  \citenamefont {Ichiguchi}, \citenamefont {Drew},\ and\ \citenamefont
  {Wolff}}]{Dobrowolska:1984_PRB}%
  \BibitemOpen
  \bibfield  {author} {\bibinfo {author} {\bibfnamefont {M.}~\bibnamefont
  {Dobrowolska}}, \bibinfo {author} {\bibfnamefont {A.}~\bibnamefont
  {Witowski}}, \bibinfo {author} {\bibfnamefont {J.~K.}\ \bibnamefont
  {Furdyna}}, \bibinfo {author} {\bibfnamefont {T.}~\bibnamefont {Ichiguchi}},
  \bibinfo {author} {\bibfnamefont {H.~D.}\ \bibnamefont {Drew}}, \ and\
  \bibinfo {author} {\bibfnamefont {P.~A.}\ \bibnamefont {Wolff}},\ }\href
  {\doibase 10.1103/PhysRevB.29.6652} {\bibfield  {journal} {\bibinfo
  {journal} {Phys. Rev. B}\ }\textbf {\bibinfo {volume} {29}},\ \bibinfo
  {pages} {6652} (\bibinfo {year} {1984})}\BibitemShut {NoStop}%
\bibitem [{\citenamefont {Kozuka}\ \emph {et~al.}(2013)\citenamefont {Kozuka},
  \citenamefont {Teraoka}, \citenamefont {Falson}, \citenamefont {Oiwa},
  \citenamefont {Tsukazaki}, \citenamefont {Tarucha},\ and\ \citenamefont
  {Kawasaki}}]{Kozuka:2013_PRB}%
  \BibitemOpen
  \bibfield  {author} {\bibinfo {author} {\bibfnamefont {Y.}~\bibnamefont
  {Kozuka}}, \bibinfo {author} {\bibfnamefont {S.}~\bibnamefont {Teraoka}},
  \bibinfo {author} {\bibfnamefont {J.}~\bibnamefont {Falson}}, \bibinfo
  {author} {\bibfnamefont {A.}~\bibnamefont {Oiwa}}, \bibinfo {author}
  {\bibfnamefont {A.}~\bibnamefont {Tsukazaki}}, \bibinfo {author}
  {\bibfnamefont {S.}~\bibnamefont {Tarucha}}, \ and\ \bibinfo {author}
  {\bibfnamefont {M.}~\bibnamefont {Kawasaki}},\ }\href {\doibase
  10.1103/PhysRevB.87.205411} {\bibfield  {journal} {\bibinfo  {journal} {Phys.
  Rev. B}\ }\textbf {\bibinfo {volume} {87}},\ \bibinfo {pages} {205411}
  (\bibinfo {year} {2013})}\BibitemShut {NoStop}%
\bibitem [{\citenamefont {Altshuler}\ and\ \citenamefont
  {Aronov}(1985)}]{Altshuler:1985_B}%
  \BibitemOpen
  \bibfield  {author} {\bibinfo {author} {\bibfnamefont {B.}~\bibnamefont
  {Altshuler}}\ and\ \bibinfo {author} {\bibfnamefont {A.}~\bibnamefont
  {Aronov}},\ }in\ \href {\doibase 10.1016/B978-0-444-86916-6.50007-7} {\emph
  {\bibinfo {booktitle} {Electron-Electron Interactions in Disordered
  Systems}}},\ \bibinfo {series} {Modern Problems in Condensed Matter
  Sciences}, Vol.~\bibinfo {volume} {10},\ \bibinfo {editor} {edited by\
  \bibinfo {editor} {\bibfnamefont {A.}~\bibnamefont {Efros}}\ and\ \bibinfo
  {editor} {\bibfnamefont {M.}~\bibnamefont {Pollak}}}\ (\bibinfo  {publisher}
  {Elsevier, Amsterdam},\ \bibinfo {year} {1985})\ pp.\ \bibinfo {pages} {1 --
  153}\BibitemShut {NoStop}%
\bibitem [{\citenamefont {Lee}\ and\ \citenamefont
  {Ramakrishnan}(1985)}]{Lee:1985_RMP}%
  \BibitemOpen
  \bibfield  {author} {\bibinfo {author} {\bibfnamefont {P.~A.}\ \bibnamefont
  {Lee}}\ and\ \bibinfo {author} {\bibfnamefont {T.~V.}\ \bibnamefont
  {Ramakrishnan}},\ }\href {\doibase 10.1103/RevModPhys.57.287} {\bibfield
  {journal} {\bibinfo  {journal} {Rev. Mod. Phys.}\ }\textbf {\bibinfo {volume}
  {57}},\ \bibinfo {pages} {287} (\bibinfo {year} {1985})}\BibitemShut
  {NoStop}%
\bibitem [{\citenamefont {Altshuler}\ \emph
  {et~al.}(1982{\natexlab{a}})\citenamefont {Altshuler}, \citenamefont
  {Aronov},\ and\ \citenamefont {Khmelnitsky}}]{Altshuler:1982_JPC}%
  \BibitemOpen
  \bibfield  {author} {\bibinfo {author} {\bibfnamefont {B.~L.}\ \bibnamefont
  {Altshuler}}, \bibinfo {author} {\bibfnamefont {A.~G.}\ \bibnamefont
  {Aronov}}, \ and\ \bibinfo {author} {\bibfnamefont {D.~E.}\ \bibnamefont
  {Khmelnitsky}},\ }\href {\doibase 10.1088/0022-3719/15/36/018} {\bibfield
  {journal} {\bibinfo  {journal} {J. Phys. C}\ }\textbf {\bibinfo {volume}
  {15}},\ \bibinfo {pages} {7367} (\bibinfo {year}
  {1982}{\natexlab{a}})}\BibitemShut {NoStop}%
\bibitem [{\citenamefont {Ferreira~da Silva}\ and\ \citenamefont
  {Persson}(2002)}]{daSilva:2002_JAP}%
  \BibitemOpen
  \bibfield  {author} {\bibinfo {author} {\bibfnamefont {A.}~\bibnamefont
  {Ferreira~da Silva}}\ and\ \bibinfo {author} {\bibfnamefont {C.}~\bibnamefont
  {Persson}},\ }\href {\doibase http://dx.doi.org/10.1063/1.1499202} {\bibfield
   {journal} {\bibinfo  {journal} {J. Appl. Phys}\ }\textbf {\bibinfo {volume}
  {92}},\ \bibinfo {pages} {2550} (\bibinfo {year} {2002})}\BibitemShut
  {NoStop}%
\bibitem [{\citenamefont {Kawabata}(1980{\natexlab{a}})}]{Kawabata:1980_JPSJ}%
  \BibitemOpen
  \bibfield  {author} {\bibinfo {author} {\bibfnamefont {A.}~\bibnamefont
  {Kawabata}},\ }\href {\doibase 10.1143/JPSJ.49.628} {\bibfield  {journal}
  {\bibinfo  {journal} {J. Phys. Soc. Jpn.}\ }\textbf {\bibinfo {volume}
  {49}},\ \bibinfo {pages} {628} (\bibinfo {year}
  {1980}{\natexlab{a}})}\BibitemShut {NoStop}%
\bibitem [{\citenamefont {Kawabata}(1980{\natexlab{b}})}]{Kawabata:1980_SSC}%
  \BibitemOpen
  \bibfield  {author} {\bibinfo {author} {\bibfnamefont {A.}~\bibnamefont
  {Kawabata}},\ }\href {\doibase
  http://dx.doi.org/10.1016/0038-1098(80)90644-4} {\bibfield  {journal}
  {\bibinfo  {journal} {Solid State Commun.}\ }\textbf {\bibinfo {volume}
  {34}},\ \bibinfo {pages} {431 } (\bibinfo {year}
  {1980}{\natexlab{b}})}\BibitemShut {NoStop}%
\bibitem [{\citenamefont {Casella}(1960)}]{Casella:1960_PRL}%
  \BibitemOpen
  \bibfield  {author} {\bibinfo {author} {\bibfnamefont {R.~C.}\ \bibnamefont
  {Casella}},\ }\href {\doibase 10.1103/PhysRevLett.5.371} {\bibfield
  {journal} {\bibinfo  {journal} {Phys. Rev. Lett.}\ }\textbf {\bibinfo
  {volume} {5}},\ \bibinfo {pages} {371} (\bibinfo {year} {1960})}\BibitemShut
  {NoStop}%
\bibitem [{\citenamefont {Altshuler}\ \emph
  {et~al.}(1982{\natexlab{b}})\citenamefont {Altshuler}, \citenamefont
  {Aronov}, \citenamefont {Khmelnitskii},\ and\ \citenamefont
  {Larkin}}]{Altshuler:1982_B}%
  \BibitemOpen
  \bibfield  {author} {\bibinfo {author} {\bibfnamefont {B.~L.}\ \bibnamefont
  {Altshuler}}, \bibinfo {author} {\bibfnamefont {A.~G.}\ \bibnamefont
  {Aronov}}, \bibinfo {author} {\bibfnamefont {D.~E.}\ \bibnamefont
  {Khmelnitskii}}, \ and\ \bibinfo {author} {\bibfnamefont {A.~I.}\
  \bibnamefont {Larkin}},\ }in\ \href@noop {} {\emph {\bibinfo {booktitle}
  {Quantum Theory of Solids}}},\ \bibinfo {editor} {edited by\ \bibinfo
  {editor} {\bibfnamefont {I.~M.}\ \bibnamefont {Lifshits}}}\ (\bibinfo
  {publisher} {Mir, Moscow},\ \bibinfo {year} {1982})\ p.\ \bibinfo {pages}
  {130}\BibitemShut {NoStop}%
\bibitem [{\citenamefont {Witowski}\ \emph {et~al.}(1999)\citenamefont
  {Witowski}, \citenamefont {Paku{\l}a}, \citenamefont {Baranowski},
  \citenamefont {Sadowski},\ and\ \citenamefont {Wyder}}]{Witowski:1999_APL}%
  \BibitemOpen
  \bibfield  {author} {\bibinfo {author} {\bibfnamefont {A.~M.}\ \bibnamefont
  {Witowski}}, \bibinfo {author} {\bibfnamefont {K.}~\bibnamefont {Paku{\l}a}},
  \bibinfo {author} {\bibfnamefont {J.~M.}\ \bibnamefont {Baranowski}},
  \bibinfo {author} {\bibfnamefont {M.~L.}\ \bibnamefont {Sadowski}}, \ and\
  \bibinfo {author} {\bibfnamefont {P.}~\bibnamefont {Wyder}},\ }\href
  {\doibase 10.1063/1.125567} {\bibfield  {journal} {\bibinfo  {journal} {Appl.
  Phys. Lett.}\ }\textbf {\bibinfo {volume} {75}},\ \bibinfo {pages} {4154}
  (\bibinfo {year} {1999})}\BibitemShut {NoStop}%
\bibitem [{\citenamefont {Majewski}\ \emph {et~al.}(2004)\citenamefont
  {Majewski}, \citenamefont {Birner}, \citenamefont {Trellakis}, \citenamefont
  {Sabathil},\ and\ \citenamefont {Vogl}}]{Majewski:2004_PSSc}%
  \BibitemOpen
  \bibfield  {author} {\bibinfo {author} {\bibfnamefont {J.~A.}\ \bibnamefont
  {Majewski}}, \bibinfo {author} {\bibfnamefont {S.}~\bibnamefont {Birner}},
  \bibinfo {author} {\bibfnamefont {A.}~\bibnamefont {Trellakis}}, \bibinfo
  {author} {\bibfnamefont {M.}~\bibnamefont {Sabathil}}, \ and\ \bibinfo
  {author} {\bibfnamefont {P.}~\bibnamefont {Vogl}},\ }\href {\doibase
  10.1002/pssc.200404761} {\bibfield  {journal} {\bibinfo  {journal} {Phys.
  Stat. Sol. (c)}\ }\textbf {\bibinfo {volume} {1}},\ \bibinfo {pages} {2003}
  (\bibinfo {year} {2004})}\BibitemShut {NoStop}%
\end{thebibliography}

%

\end{document}